\documentclass[12pt, onecolumn]{IEEEtranTCOM}
\usepackage[utf8]{inputenc}
\usepackage{mathtools}
\usepackage{amssymb}
\usepackage{lmodern}
\usepackage{newtxtext}
\usepackage{graphicx}
\usepackage{hyperref}
\usepackage{url}
\usepackage{bbm}
\usepackage[mathscr]{euscript}
\usepackage{setspace}
\usepackage{cite}
\DeclareMathOperator{\E}{\mathbb{E}}
\DeclareMathOperator{\Rc}{\mathcal{R}}
\DeclareMathOperator{\B}{\mathcal{B}}
\DeclareMathOperator{\R}{\mathscr{R}}
\allowdisplaybreaks
\makeatletter
\def\endthebibliography{%
	\def\@noitemerr{\@latex@warning{Empty `thebibliography' environment}}%
	\endlist
}
\makeatother

\begin{document}
	\title{{Adaptive Acquisition Schemes for Photon-Limited Free-Space  Optical Communications}}
	\author{Muhammad~Salman~Bashir,~\IEEEmembership{Member,~IEEE,}
		and~Mohamed-Slim~Alouini,~\IEEEmembership{Fellow,~IEEE}
		\thanks{This work is supported by Office of Sponsored Research (OSR) at King Abdullah University of Science and Technology (KAUST). \newline
			M.~S.~Bashir and M.~-S.~Alouini  are with the King Abdullah University of Science and Technology (KAUST), Thuwal 23955-6900, Kingdom of Saudi Arabia.  e-mail: (muhammad.bashir@fulbrightmail.org, slim.alouini@kaust.edu.sa).}
		\thanks{}
		\thanks{}}
	
	\markboth{}%
	{}
	
	\markboth{}%
	{}
	%

	\maketitle
\begin{abstract}
	Acquisition and tracking systems form an important component of free-space optical communications due to directional nature of the optical signal. Acquisition subsystems are needed in order to search and locate the receiver terminal in an uncertainty/search region with very narrow laser beams. In this paper, we have proposed and analyzed two adaptive  search schemes for acquisition systems that perform better---for the low probability of detection---than the spiral scanning approach. The first of these schemes, the adaptive spiral search, provides a better acquisition time performance by dividing the search region into a number of smaller subregions, and prioritizing search  in regions of higher probability mass. The second technique---the shotgun approach---searches the region in a random manner by sampling the search region according to a Gaussian distribution. The adaptive spiral scheme outperforms the shotgun approach in terms of acquisition time, especially if the number of search subregions is large enough. However, a higher pointing accuracy is required by the adaptive spiral search in order to search the region precisely. On the other hand, the shotgun scanning approach does not require such stringent pointing accuracy.
\end{abstract}

\IEEEpeerreviewmaketitle

\begin{figure}
    \centering
    \includegraphics[scale=0.7]{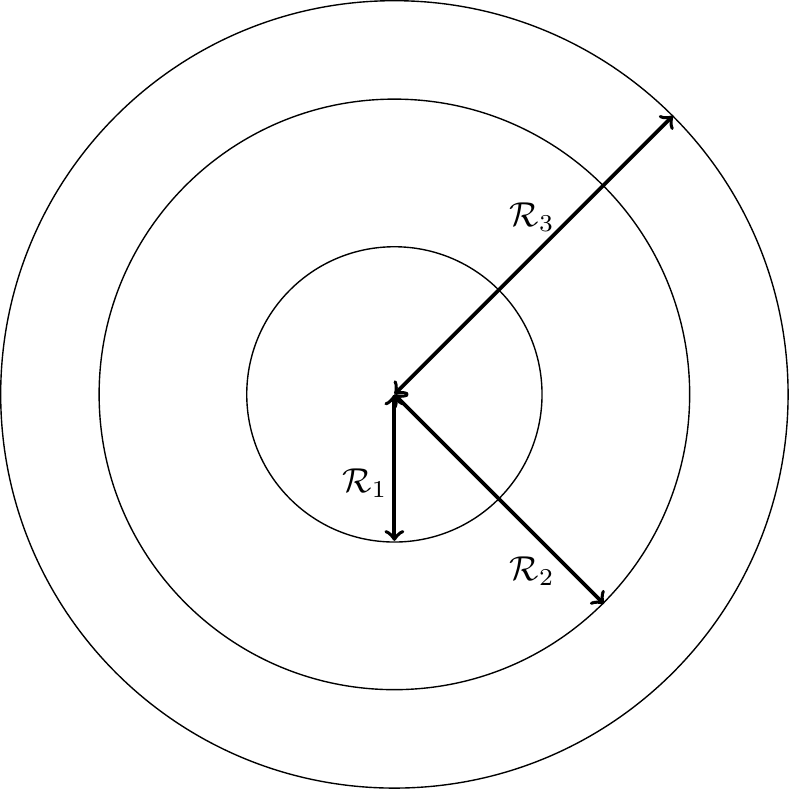}
    \caption{Acquisition system with $N = 3$. The radius of the uncertainty region is $\mathcal{R}_3$.}
    \label{fig_acq}
\end{figure}
\section{Introduction}
Free-space optical (FSO) communications is a promising technique that can provide high data-rates for the next generation of wireless communication systems.  Because of the availability of large chunks of unregulated spectrum available in the optical domain, high-speed data communications can be achieved with FSO systems. These systems have typically been used in deep space communications where the long link distances dictate that the transmitted energy be focused to achieve a small angle of divergence. However, more recently, big internet corporations---such as Facebook and Google---are employing FSO in the backhaul network in order to provide connectivity to regions of the world that still lack internet access \cite{Kaymak}.

Because of the narrow beamwidth associated with the optical signal---as is the case for any ``directional'' communication system, such as Terahertz and millimeter wave systems---acquisition and tracking subsystems are needed in order to establish and maintain the link between the transmitter and the receiver terminals, respectively.  \emph{Acquisition} is a process whereby two terminals obtain each others location in order to effectively communicate in a directional communications setting. 

\subsection{Motivation}
In this paper, we have analyzed the adaptive acquisition schemes for low probability of detection/photon-limited optical channels. For a small detection probability, these adaptive algorithms provide a significantly better performance than the nonadaptive search schemes used in state-of-the-art acquisition systems. An example of a nonadaptive scheme is the spiral search that is argued to be optimal for a Rayleigh distributed receiver location in the uncertainty region, and outperforms other scanning approaches when the probability of detection is high \cite{XinLi, Bashir6}. However, for photon-limited channels that incur a small probability of detection, this scheme does not perform as well (as will be shown later in this study). 

Photon-limited channels exist in deep space communications where the long link distances result in a significant reduction of received signal photons \cite{Griffiths_MDPI:2018}. Additionally, such channels also exist in terrestrial FSO where the presence of fog or clouds results in a significant attenuation of transmitted energy. Because of low numbers of received signal/receiver noise photons \cite{Bell_AO:2000}\footnote{With the help of cryogenic receivers, the number of noise photons can be reduced significantly \cite{Griffiths_MDPI:2018}.}, the probability of detection for a \emph{Pulse Position Modulation} (PPM) or \emph{On-Off Keying} (OOK) receiver can take a serious hit. This also affects the acquisition performance since successful acquisition depends on detection probability of the transmitted pulse at the receiver. For the spiral scan,  such low photon-rate channels will lead to several scans of the uncertainty region before the terminal is discovered. This wastes both time and energy during the acquisition stage.

In addition to low photon rates, the probability of detection also suffers from a desire to achieve a low probability of false alarm during the acquisition stage. A reasonably low probability of false alarm is needed so that we do not ``misacquire'' the terminal: that is, the transmitter mistakenly decides that the receiver has been located in the uncertainty region, and begins to transmit data in the ``wrong'' direction. This misacquisition wastes energy and time, and results in restarting the acquisition process after the misacquisition event is detected. Therefore, during the detection process in the acquisition stage, we have to set the threshold high enough in order to set the probability of false alarm reasonably low. However, setting the threshold higher than usual also results in a lower probability of detection (please see the arguments in Section~\ref{Detect} regarding the threshold selection). After the acquisition stage is completed successfully, the threshold can be lowered in order to increase the probability of detection (or minimize the probability of error) for the purpose of decoding data symbols.

The photon counting channel is modeled by a \emph{Poisson Point Process} (PPP). The studies \cite{Snyder} and \cite{Streit} provide an elegant treatment of the theory of PPP's.

\subsection{Related Literature Review}
The focus of this study is on attacking the acquisition problem in FSO purely from a signal processing/probabilistic perspective. Therefore, we will only review the papers that adopt a similar approach to the acquisition problem. In this regard, we were able to find three major studies on acquisition. The first article \cite{Wang} is focused on  realizing secure acquisition between two mobile terminals. The idea is to use a double-loop raster scan so that the reception of the signal and the verification of identities through a IV code can be carried out in rapid succession. They have proposed an array of detectors at the receiver that acts both as a bearing/data symbol detector. The acquisition time is optimized in terms of signal-to-noise ratio and beam divergence among other parameters.  

The authors in \cite{XinLi} optimize the acquisition time as a function of the uncertainty sphere angle. Instead of scanning the entire uncertainty region, their idea is to scan a subregion of the uncertainty sphere that contains the highest probability mass. This is done in order to save time. The acquisition is carried out for a mobile satellite scenario, whose location coordinates at a certain point in time---obtained through ephemeris data---is designated as the center of the uncertainty sphere. The spiral scanning technique is used to locate the satellite. Instead of searching the whole sphere (three standard deviations for a Gaussian sphere), they search a fraction of the region (which is 1.3 times the standard deviation). If the satellite is missed in one search, the hope is that it will be located in the next search, and so on\footnote{The point in space for the next search, which will form the center of the new uncertainty sphere, is obtained from the ephemeris data. }.

The authors in \cite{Bashir6} describe the signal acquisition technique for a stationary receiver that employs an array of small detectors. They conclude that an array of detectors minimizes the acquisition time as compared to one single detector of similar area as an array. They also consider the possibility of multiple scans of the uncertainty region in case the receiver is not acquired after a given scan. An upper bound on mean acquisition time is optimized with respect to beam radius, and the complementary cumulative distribution function of the upper bound is computed in closed-form.

There is another body of work that discusses improvement in acquisition/tracking performance by offering hardware-based solutions. In this regard, we will cite a few important studies. The authors in \cite{Deng} propose to improve tracking performance with the help of camera sensors that direct the movement of control moment gyroscopes (CMG) in order to control a bifocal relay mirror spacecraft assembly. The main application of their work is to minimize the jitter/vibrations in the beam position using CMG's and fine tracking using fast steering mirrors. The work \cite{Kim} adopts gimbal less Micro-Electro-Mechanical Systems (MEMS) micro-mirrors for fast tracking of the time-varying beam position. The authors in \cite{Rzasa} examine the acquisition performance of a gimbal based pointing system in an experimental setting that utilizes spiral techniques for searching the uncertainty sphere. 

For literature on pointing error in free-space optical communications, the readers are referred to some recent work, such as \cite{Ansari, Zedini, Quwaiee, Issaid, Farid}. In order to comprehend tracking with an array of detectors, \cite{Bashir1, Bashir2, Bashir7} are helpful. 

\subsection{Contributions of This Study} \label{cont}
In this paper, we have devised two adaptive acquisition schemes for photon-limited FSO channels. In the first part of the paper, we propose an adaptive acquisition scheme that divides the uncertainty region into a number of smaller subregions, and the subregions that correspond to the higher probability mass of the receiver's location are searched more frequently than the others. The intuition behind this scheme is the following argument: If the receiver is not discovered during the search of a subregion that has a higher probability mass attached to it, then there is a higher chance that we missed the receiver due to low probability of detection, and we can achieve better performance if we rescan this particular subregion a few times before we move on to explore subregions of lower probability mass. The scanning is done by search along a spiral, and significantly better performance can be obtained by optimizing the volumes of the subregions. We call this scheme the \emph{adaptive spiral search} technique.

In the second part, we propose the \emph{shotgun} method which is a randomized acquisition scheme. In shotgun approach, the uncertainty region is scanned at locations that are sampled from a Gaussian distribution (also called the \emph{firing} distribution). By choosing the suitable variance of the firing distribution, the acquisition time can be minimized.

For a low probability of detection, both these schemes provide a better acquisition time performance than the spiral search scheme given in \cite{Bashir6} and \cite{XinLi}. As we will see later in this study, the adaptive spiral search technique significantly outperforms the shotgun approach. However, the cost we pay is the requirement to meet ultra precise pointing of the beam on the spiral during scanning process. In contrast, the shotgun approach can do without stringent requirements on pointing accuracy.

\subsection{Organization of This Paper}
This paper is organized as follows. Section~\ref{UR} defines the uncertainty region, and the location statistics of the receiver in the uncertainty region. Section~\ref{Detect} deals with the derivation of the probability of missed detection and false alarm for the acquisition process. Section~\ref{ASSP} discusses introductory material pertaining to the adaptive spiral search technique. Section~\ref{ASS} explains the adaptive spiral search scheme, and walks the reader through the derivation of mean and complementary distribution function of acquisition time.  The optimization of the acquisition time as a function of the radii of subregions is discussed in the same section. Section~\ref{SG} examines the shotgun approach.  Section~\ref{Comp} compares the two acquisition approaches, and Section~\ref{Conc} summarizes the conclusions of this study.

\section{Uncertainty Region and Scanning Technique} \label{UR}
The \emph{uncertainty region}---or \emph{uncertainty sphere}, or the \emph{search region}---is a volume in space that is scanned by the initiator/transmitter terminal to locate the receiver terminal in order to establish a communications link. As discussed in detail in \cite{Bashir6}, the errors in the measurements of localization systems, and the errors in the pointing assembly of the transmitter, contribute to the volume of the uncertainty region: the larger the error variance, the greater the volume the transmitter has to scan in order to successfully complete the acquisition stage. 

The error in two dimensions in the uncertainty region is modeled by a two dimensional Gaussian distribution. If the error in each dimension is assumed to be independent with equal variance, the resulting distribution is a circularly symmetric Gaussian distribution, and the distance from the center is modeled as a Rayleigh distributed random variable. For the spiral scan technique, the acquisition time in this case becomes tractable to analyze since the time it takes to start from the center of the uncertainty region to the point where the receiver is located is modeled approximately by an exponential distribution for the successful detection scenario. However, as discussed in \cite{Bashir6}, the uncertainty region, in general, is represented by a general (elliptical) Gaussian distribution in two dimensions (correlated Gaussian errors in two dimensions with unequal variance). Nevertheless, as argued in \cite{Bashir6}, if the general error covariance matrix is known, any elliptical uncertainty region can be transformed to a circular uncertainty region by using an appropriate linear transformation, and the probability distribution of the acquisition time in the circular uncertainty region case is the same as the acquisition time distribution in the elliptical case.
\begin{figure}
	\centering
	\includegraphics[scale=0.9]{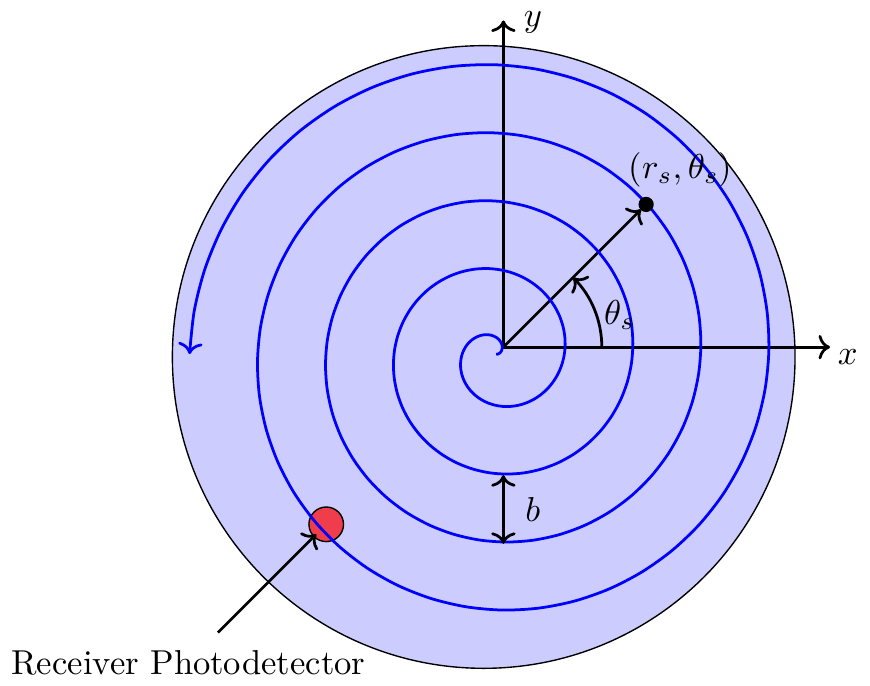}
	\caption{The transmitter scans the uncertainty sphere along the Archimedean spiral (shown in blue) in order to search the receiver. The distance between consecutive rotations $b$, and the step size on the spiral, is chosen according to the beam width of the scanning beam. The location at any point on the spiral is represented by $(r_s, \theta_s)$.} \label{ur}
\end{figure}

For a circular uncertainty region, the Archimedean spiral search technique provides an optimal performance in terms of acquisition time \cite{XinLi}. Since the spiral search scans the contours of higher probability mass first as opposed to contours of lower probability, it is easy to see---at least intuitively---as to why the spiral search will perform better than other search techniques for a circular uncertainty region. Fig.~\ref{ur} shows an example of a circular uncertainty sphere.  For a general (elliptical) uncertainty sphere, the spiral scan method will be replaced by a similar technique that starts the scan from the center, and then moves outward along elliptical contours of Gaussian distribution.

Let us say that there are two terminals---Terminal~A and Terminal~B---that want to set up a communication link with each other. In order to initiate the spiral scan, let us assume that Terminal~A will begin by pointing towards the center of the uncertainty region, transmit a pulse, and then listen for any feedback information from Terminal~B. If Terminal~B detects the pulse, it will send a signal back to Terminal~A on a low data-rate radio frequency (RF) feedback channel to confirm that the signal has been acquired. Otherwise, A will point to the next point on the spiral and transmit, and the process repeats itself until Terminal~B has been found. The time that Terminal~A waits before transmitting the next pulse is known as the \emph{dwell time}, and this time interval takes into account factors, such as receiver processing time and the round-trip-delay time. For more details in this regard, the reader is referred to \cite{Bashir6}. Once, Terminal~A discovers Terminal~B, Terminal~B starts the same process in order to locate Terminal~A. However, since Terminal~B now has information about Terminal~A's angle-of-arrival, the search region to locate Terminal~A will be much smaller. Thus, the total acquisition time is approximately the time that Terminal~A requires in order to locate Terminal~B. 

For a randomized search scheme such as the shotgun approach, the receiver location process  is similar, except that the initiator/transmitter fires pulses at random locations inside the uncertainty region instead of moving precisely along a spiral.

The receiver (Terminal~B) detects the pulse by executing a hypothesis test. The probability of detection and the probability of false alarm at the receiver is the subject of discussion in the next section.

\section{Probability of Detection/False Alarm} \label{Detect}

\subsection{Probability of Detection}
The system will decide whether the receiver is detected at a given point in the uncertainty region by carrying out the following hypothesis test:
\begin{align}
\frac{p(Z|H_1) p(H_1)}{p(Z|H_0) p(H_0)} \underset{H_0}{\overset{H_1}{\lessgtr} } \gamma, 
\end{align}
where $Z$ is the (random) photon count generated in the optical detector during an observation period, $\gamma$ is a (positive) threshold, and $H_1$ is the hypothesis that the terminal is present at a given point in $S(\Rc)$, and $H_0$ is the hypothesis that the receiver is not present. The probability of detection for a \emph{maximum a posteriori probability} (MAP) detector is 
\begin{align}
P_D = P\left( \left\{\frac{p(Z|H_1) }{p(Z|H_0) } > \gamma_0 \right\} \right), \label{PD}
\end{align}
where $p(Z|H_1) \coloneqq \frac{e^{-(\lambda_s + \lambda_n) A T} ((\lambda_s + \lambda_n )A T)^Z}{Z!},$
$p(Z|H_0) \coloneqq \frac{e^{-\lambda_n A T} (\lambda_n A T)^Z}{Z!}$,
$P(H_1) = \frac{r}{\sigma^2}e^{-\frac{r^2}{2\sigma^2}}, r \geq 0,$
and $P(H_0) = 1 - P(H_1).$ Additionally, $\gamma_0 \coloneqq \gamma \frac{p(H_0)}{p(H_1)}$. The number $r$ is the distance from the center of the uncertainty region. The quantity $(\lambda_s + \lambda_n)AT$ refers to the mean photon count for the signal plus noise ($H_1$) hypotheis, and $\lambda_n$ refers to the mean photon count for the noise only ($H_0$) hypothesis.   The quantity $A$ is the area of the detector, and $T$ represents an observation interval. The constant $\gamma$ is an appropriate threshold chosen for some fixed \emph{probability of false alarm}, $P_{FA}$. Specifically, 
$ P_{FA} = P\left( \left\{\frac{p(Z|H_1) }{p(Z|H_0) } > \gamma_0 \right\} \right),$
where $Z$ is Poisson with mean $\lambda_n AT$. 

The probability of detection $P_D$ is a function of the signal power $\lambda_s AT$. The intensity of light, $\lambda_s$, that is impinging on the detector is usually assumed to be Gaussian distributed in two dimensions. In order to simplify the analysis, we approximate the Gaussian function with a cylinder function: the intensity is uniform over a circular region of radius $\rho$, and is zero elsewhere. 
Thus, for a constant transmitted signal power $P_s$, $\lambda_s$ should drop as $\rho$ is enlarged since $P_s = \lambda_s  \pi \rho^2$, where $P_s$ is the transmitted signal power. Thus, $p(Z|H_1)$  becomes
\begin{align}
p(Z|H_1) \coloneqq \frac{\exp\left(-\left(\frac{P_s}{ \pi\rho^2} + \lambda_n \right) A T \right)   \left( \left( \frac{P_s}{\pi \rho^2} + \lambda_n \right) A T \right)^Z}{Z!}.
\end{align}
This goes to show that $P_D$ is a function of $\rho$ through the dependence of $p(Z|H_1)$ on $\rho$. 

The probability of detection, $P_D$, can be simplified analytically. The log-likelihood ratio can be written as
\begin{align}
\mathcal{L}(Z) & =   \ln p(Z|H_1)   - \ln p(Z|H_0)  = -\left(\frac{P_s}{ \pi\rho^2} + \lambda_n \right) A T  + Z \ln \left(\left(\frac{P_s}{ \pi\rho^2} + \lambda_n \right) A T  \right)  + \lambda_n A T - Z \ln(\lambda_n A T).
\end{align}
Therefore, 
\begin{align}
P_D &= P(\{ \mathcal{L}(Z) > \ln(\gamma)\}) = P\left( \left\{ Z > \gamma_0  \right\} \right) = 1 - P(\{ Z \leq \gamma_0 \}) \nonumber \\
&= 1 - \sum_{z=0}^{\lfloor \gamma_0 \rfloor} \frac{\exp\left(-\left(\frac{P_s}{ \pi\rho^2} + \lambda_n \right) A T \right)   \left( \left( \frac{P_s}{\pi \rho^2} + \lambda_n \right) A T \right)^z}{z!}= 1 - Q\left(\lfloor \gamma_0 + 1\rfloor,  \left(\frac{P_s}{ \pi\rho^2} + \lambda_n \right) A T  \right). \label{PD1}
\end{align}
The function $Q(x, y)$ is known as the \emph{regularized Gamma function} and is defined as 
$
Q(x, y) \triangleq \frac{\Gamma(x,y)}{\Gamma(x)}, \label{rgamma}
$
where $\Gamma(x, y)$ is the \emph{upper incomplete Gamma function}:
$
\Gamma(x, y) \triangleq   \int_{y}^{\infty} t^{x-1} e^{-t}\, dt,
$
and $ \Gamma(x) \triangleq \int_{0}^{\infty} t^{x-1} e^{-t}\, dt$. 

Finally, \begin{align}
P_{FA} = 1 - Q\left(\lfloor \gamma_0 + 1\rfloor,   \lambda_n  A T  \right). \label{PFA}
\end{align}

\section{Adaptive Spiral Search: Preliminaries}\label{ASSP}
In the adaptive spiral search, the uncertainty region is divided into $N$ smaller regions or subregions, $S(\Rc_i)$ for $i= 0, \dotsc, N$, where $S(\Rc_i)$ is a sphere, centered at the origin, with radius $\Rc_i$, and $\Rc_0 < \Rc_1 < \dotsm < \Rc_N$, which implies that $S(\Rc_0) \subset S(\Rc_1) \subset \dotsm \subset S(\Rc_N)$. Additionally $\Rc_0 \coloneqq 0$, $S(\Rc_0) = \phi$, $\Rc \coloneqq \Rc_N$, and $S(\Rc_N)$ corresponds to the total uncertainty region. Fig.~\ref{fig_acq} shows that the uncertainty region $S(\Rc)$ divided into three subregions.

In order to locate the receiver, the transmitter begins scanning from the origin (center of $S(\Rc)$) and finishes scanning $S(\Rc_1)$. If the receiver is not detected in this attempt, the transmitter initiates the second subscan by starting from the origin, and this time ends the scanning process when it has finished searching $S(\Rc_2)$. Thus, when the transmitter finishes searching $S(\Rc_2)$, it scans $S(\Rc_1)$ one more time and then scans the annular ring $S(\Rc_2)-S(\Rc_1)$. Hence, $S(\Rc_1)$ gets scanned a total of two times and $S(\Rc_2)-S(\Rc_1)$ is searched once  when the transmitter ends its second subscan. In a similar vein, when the transmitter has finished scanning $\Rc_N$, region $S(\Rc_1)$ gets scanned $N$ times, region $S(\Rc_2)$ gets scanned $N-1$ times, and so on. We should make it clear here that the term \emph{subscan} is used to indicate a search attempt corresponding to a particular $S(\Rc_k), k = 1, \dotsc, N$, and we say that a (single) \emph{scan} has taken place when $S(\Rc_1)$ has been searched $N$ times, $S(\Rc_2)$ is searched $N-1$ times, and so on, until the region $S(\Rc_N)$ is searched once.  If the receiver is not located during the first scan, the whole process is repeated until the time the receiver is located.

Let us now consider the smallest sphere $S(\Rc_1)$ in the adaptive spiral scan. The time taken to scan this sphere is approximately $ T_d \frac{\Rc_1^2}{\rho^2}$, where $T_d$ is the dwell time. In this case, 
\begin{align}
     P(E_1) = P( E_{S_1}\cap E_{D_1}),
\end{align}
where  $E_{S_1}$ is the event that receiver is present inside the sphere $S(\Rc_1)$, and $E_{D_1}$ is the event that receiver is detected in $S(\Rc_1)$. Moreover, $E_1$ is the event that the receiver is detected during the first attempt/subscan. It follows that, 
\begin{align} 
P(E_1) = P( E_{S_1} \cap E_{D_1} ) = \frac{P(E_{S_1} \cap E_{D_1}) } {P(E_{S_1})} P(E_{S_1}) = P(E_{D_1}|E_{S_1}) P(E_{S_1}). 
\end{align} 

In a similar fashion,
\begin{align}
    E_{k} =  A_1 \cup A_2 \cup \dotsm \cup A_k,
\end{align}
where $E_k$ is the event that the receiver is detected during the $k$th attempt/subscan. Let $E_{S_k}$ is the event that receiver is present inside the sphere $S(\Rc_k)$, and $E_{D_k}$ be the event that receiver is detected in $S(\Rc_k)$. 
The set $A_i$, for $i = 1, \dotsc, k$, is the event that the receiver lies in the set  $\left(S(\Rc_i) - S(\Rc_{i-1}) \right)$, and is not detected in $(k-i)$ attempts, and detected in one attempt. 

The set $S(\Rc_i)- S(\Rc_{i-1})$ represents the difference set. It represents the annular ring formed by the difference of two concentric spheres: $S(\Rc_i)$ and $S(\Rc_{i-1})$. The reader may note that the sets $A_{i-1}\cap A_{i} = \phi$ since $\left(S(\Rc_i)-S(\Rc_{i-1})\right) \cap S(\Rc_{i-1}) = \phi.$ Thus, 
\begin{align}
P(E_k) = \sum_{i=1}^k P(A_i). \label{Ek1}
\end{align}  

We assume that the uncertainty in the location of the receiver is modeled by zero-mean, i.i.d, Gaussian random variables with variance $\sigma^2$. Let $E_{S_k}$ be the event that the receiver is present in the sphere $S(\Rc_k) $. Then, we have that, 
\begin{align}
P(E_{S_k}) = \int_0^{\mathcal{R}_k} f_R(r)\, dr, \quad k = 1, \dotsc, N,
\end{align}
where $f_R(r) \coloneqq \frac{r}{\sigma^2} e^{-\frac{r^2}{\sigma^2}}, r \geq 0$.

From \eqref{Ek1},  
\begin{align}
    P(E_k)& = \sum_{i=1}^{k} P\left(E_{S_i}-E_{S_{i-1}} \left| \bigcap_{j=1}^{k-1} E_j^C \right. \right) (1-P_D)^{k-i} P_D \nonumber \\
    &= \sum_{i=1}^{k} \left[ P\left(E_{S_i} \left| \bigcap_{j=1}^{k-1} E_j^C \right. \right) - P\left(E_{S_{i-1}} \left| \bigcap_{j=1}^{k-1} E_j^C \right) \right. \right] (1-P_D)^{k-i} P_D,  \label{Ek}
\end{align}
where $P_D \coloneqq P(E_{D_1}| E_{S_1}) = \dotsm = P(E_{D_{k}} | E_{S_{k}})$. The probability of detection, $P_D$, is derived in Section~\ref{Detect}. It is shown in Appendix~\ref{A} that for a small $P_D$,  $P \left( E_{S_i} \left| \bigcap_{j=1}^{k-1} E_j^C \right. \right) \approx P(E_{S_i})$ for $i = 1, 2, \dotsc, k$. This means that for a small $P_D$, the observation that the receiver has not been located in the previous $k-1$ attempts does not alter the receiver's location distribution for the $k$th attempt. This is an important result that simplifies the analysis considerably. Therefore,   
\begin{align}
P(E_k) \approx \sum_{i=1}^k \left[ P\left(E_{S_i} \right) - P\left(E_{S_{i-1}} \right)  \right] (1-P_D)^{k-i} P_D.  \label{Ek2}
\end{align}

Let the event $F$ denote the event that given the receiver is present in the uncertainty region $S(\Rc)$, the acquisition system fails to  locate the receiver during one full scan of $S(\Rc)$ through the adaptive scheme. Then
\begin{align}
P(F) = \sum_{k=0}^{N-1} P(E_{S_{k+1}} - E_{S_{k}}) (1 - P_D)^{N-k},  \label{PF} 
\end{align}
where $E_{S_0} \coloneqq \phi$, the empty set. We note that for the nondaptive acquisition scheme, 
\begin{align}
P(F) = 1 - P_D.
\end{align}

\section{ Adaptive Spiral Search} \label{ASS}
\subsection{Single Scan of $S(\mathcal{R})$}
Due to low probability of detection, we may have to carry out a number of scans before the receiver is discovered in the uncertainty region. In this section, we are focused on the amount of time spent in the (successful or final) scan in order to locate the receiver, and we represent it by the random variable $V$. Then, $V$ is a ``mixed'' random variable, and is defined as
\begin{align}
    V \coloneqq  Y + X,
\end{align}
where $X$ is the random amount of time it takes for the system to detect the receiver during a ``successful'' subscan, and $Y$ represents the distribution of time that is ``wasted'' in unsuccessful subscans of the final scan. It can be seen that the value or distribution of $X$ will depend on area of the region in which the successful detection of the receiver takes place. Thus, given that the receiver is detected during the $k$th subscan, it is shown in the Appendix~\ref{B} that the conditional pdf of $X$ is represented by a \emph{truncated exponential} distribution:
\begin{align}
    f_X(x|E_k) = \frac{1}{\eta_k}\alpha \exp\left( -\alpha x   \right) \cdot \mathbbm{1}_{[0, T_d\Rc_k^2/\rho^2]}(x),
    \end{align}
where $\mathbbm{1}_A(x)$ is the indicator function over some (measurable) set $A$, and $\eta_k$ is the normalization constant given by 
\begin{align}
    \eta_k \coloneqq \int_0^{\frac{T_d\Rc_k^2}{\rho^2}} \alpha e^{-\alpha x} \, dx = 1 - e^{-\alpha T_d \frac{\Rc_k^2}{\rho^2} },
\end{align}
where $\alpha \coloneqq \frac{\rho^2}{2 T_d \sigma^2}$  \cite{Bashir6}. 

Before we define the distribution of $Y$, let us define $\R_k \coloneqq \sum_{i=1}^k \Rc_i^2$, $k = 1, \dotsc, N$. Then, the random variable $Y$ has a discrete distribution, and takes on the following values:
    $Y = T_d\frac{ \R_1}{\rho^2}$ when the receiver is detected in region $S(\mathcal{R}_2)$, $Y = T_d \frac{\R_2}{\rho^2}$  when the receiver is detected in region $S(\mathcal{R}_3)$, and $Y = T_D \frac{\R_{N-1}}{\rho^2}$ when the receiver is detected in $S(\mathcal{R})$. Finally, if the acquisition process fails in $S(\mathcal{R})$, then $Y = T_d\frac{\R_N}{\rho^2}$. 
More compactly, 
\begin{align}
f_Y(y) = \sum_{k=0}^{N-1} P(E_{k+1})\delta\left( y - T_d\frac{\R_k}{\rho^2} \right)  + P(F) \delta\left( y - T_d \frac{\R_N}{\rho^2} \right), \quad y > 0, \label{Y}
\end{align}
where $\delta(x)$ is the \emph{Dirac Delta Function}, and $\R_0 \coloneqq 0.$

When the next subscan starts, the prior information about the location of the receiver inside the uncertainty region remains unchanged. This is true because of the low probability of detection argument as discussed in Section~\ref{ASSP}.  In other words, the value of $Y$ at any point does not give us any additional information about $X$. Thus $Y$ and $X$ are treated as independent random variables. For this scenario, 
\begin{align}
    f_V(v) = f_Y * f_X (v),
\end{align}
where $*$ represents the convolution operator.  Therefore, we have that,
\begin{align}
    f_V(v) &=   \sum_{k=0}^{N-1} P(E_{k+1}) \frac{\alpha}{\eta_{k+1}} e^{ -\alpha \left(v - T_d\frac{\R_k}{\rho^2} \right) } \cdot \mathbbm{1}_{\left[T_d \R_k/\rho^2, T_d(\R_k + \Rc_{k+1}^2)/\rho^2\right)}(v)+  P(F) \delta\left( v - T_d \frac{\R_N}{\rho^2} \right)
\end{align}
for $0 \leq v \leq T_d\frac{\R_n}{\rho^2}$. We note that $\R_k + \Rc_{k+1}^2 = \R_{k+1}$. Additionally, for the sake of brevity, let us denote the factor $T_d\frac{\R_k}{\rho^2}$ by $\beta_k$ for $k = 0, \dotsc, N$. Then, 
\begin{align}
    f_V(v) &=   \sum_{k=0}^{N-1} P(E_{k+1}) \frac{\alpha}{\eta_{k+1}} e^{ -\alpha \left(v - \beta_k \right) } \cdot \mathbbm{1}_{\left[\beta_k, \beta_{k+1} )/\rho^2\right)}(v)+  P(F) \delta\left( v - \beta_N \right)
\end{align}

\subsection{Multiple Scans of $S(\mathcal{R})$}
In case the event $F$ occurs, we will have to repeat the whole scanning process. In this regard, let us define the total acquisition time as  
\begin{align}
    T = W + V'. \label{T}
\end{align}
The random variable $W$ represents the time it takes to complete multiple scans of the uncertainty region $S(\mathcal{R})$ with the adaptive scheme.  It is given by
\begin{align}
    W \coloneqq U \beta_N, \label{W}
\end{align}
where $\beta_N \coloneqq  T_d \frac{\R_N}{\rho^2}$, and $U$ is a \emph{geometric} random variable with success probability $p \coloneqq P(F)$. The (discrete) distribution of $W$ is as follows:
\begin{align}
    f_W(w) = (1-p)\sum_{i=0}^\infty p^{i}\delta(w - i \beta_N). \label{fW}
\end{align}
The random variable $V'$ is a modified version of random variable $V$, since $V'$ represents the amount of time taken in the final scan of the uncertainty region given that the successful detection of the receiver occurs in this particular scan, when the previous $W$ scans have failed to locate the receiver. Thus, there is no possibility of a ``failure'' in the final scan. Therefore, the distribution of $V'$ is the same as the distribution of $V$ given that the detection event, $D$, will occur in the final scan. That is, $f_{V'}(v) = f_V(v|D)$, where we obtain $f_V(v|D)$ in a straightforward manner by using \emph{law of total probability}:
\begin{align}
&f_V(v) =   f_V(v|D)P(D) + f_V(v|D^C)P(D^C) \implies f_V(v|D) = \frac{f_V(v) - f_V(v|D^C)P(D^C)}{P(D)}
\end{align}
Since $f_V(v|D^C) = \delta(v-\beta_N)$ and $P(D) \coloneqq 1 - p$, and $P(D^C)= p$, 
\begin{align}
    f_{V'}(v) = f_V(v|D)= \frac{1}{1-p} \sum_{k=0}^{N-1} P(E_{k+1}) \frac{\alpha}{\eta_{k+1}} e^{ -\alpha \left(v - \beta_k \right) } \cdot \mathbbm{1}_{\left[\beta_k, \beta_{k+1} \right)}(v), \quad 0 \leq v < \beta_N.
\end{align}

From \eqref{T}, $f_{T}(t) = f_{W} * f_{V'} (t)$.  Then,
\begin{align}
f_T(t) &=  \sum_{i=0}^\infty p^i   \sum_{k=0}^{N-1} P(E_{k+1}) \frac{\alpha}{\eta_{k+1}} e^{-\alpha \left(t - i\beta_N - \beta_k \right) }\cdot \mathbbm{1}_{[i\beta_N + \beta_k, i\beta_N + \beta_{k+1})}(t).  
\end{align}
\subsection{ Expected Value  and Complementary Cumulative Distribution Function of $T$}
\subsubsection{Expected Value}
The expected value of  acquisition time is
\begin{align}
    &\E[T] = \int_{-\infty}^{\infty} t f_T(t)\, dt = \sum_{i=0}^\infty p^i   \sum_{k=0}^{N-1} P(E_{k+1})\int_{-\infty}^{\infty} t \frac{\alpha}{\eta_{k+1}} e^{-\alpha \left(t - i\beta_N - \beta_k \right) }\cdot \mathbbm{1}_{[i\beta_N + \beta_k, i\beta_N + \beta_{k+1})}(t)\, dt  \\
    &= \sum_{i=0}^\infty  p^i  \sum_{k=0}^{N-1}  \frac{P(E_{k+1})}{\eta_{k+1}} \left( i\beta_N + \beta_k + \frac{1}{\alpha}   - \left( i\beta_N + \beta_{k+1} +\frac{1}{\alpha} \right)e^{ -\alpha \left(\beta_{k+1} - \beta_k \right) } \right) 
    \end{align}
    After a few more simplifications, we have that 
\begin{align}
    \E[T]& = \sum_{k=0}^{N-1}\frac{P(E_{k+1})}{\eta_{k+1}} \left( \sum_{i=0}^\infty i p^i  \beta_N \left( 1 - e^{-\alpha(\beta_{k+1}-\beta_k)} \right) + \sum_{i=0}^\infty p^i \left( \beta_k - \beta_{k+1}e^{-\alpha\left( \beta_{k+1}-\beta_k \right)} + \frac{1}{\alpha} \left( 1 - e^{-\alpha \left( \beta_{k+1} -\beta_k \right)} \right) \right)   \right).
\end{align}
The quantity  $\sum_{i=0}^\infty i p^i = \frac{p}{(1-p)^2}$. Thus, 
\begin{align}
    &\E[T] =   \sum_{k=0}^{N-1}\frac{P(E_{k+1})}{\eta_{k+1}} \left( \beta_N\left( 1 - e^{-\alpha \left( \beta_{k+1} - \beta_k \right)} \right) \frac{p}{(1-p)^2} + \frac{\beta_k - \beta_{k+1}e^{-\alpha \left( \beta_{k+1} - \beta_k \right)} + \frac{1}{\alpha} \left( 1 - e^{-\alpha \left( \beta_{k+1} - \beta_k \right)} \right)   }{1-p} \right).
\end{align}

It should be noted that $P(E_{k+1})$, $p$, and $\beta$, are all  functions of $\Rc_i$'s (or $r_i$'s). Specifically, using \eqref{Ek} and \eqref{PD1},
\begin{align}
   P( E_{k+1}) = \left(1 - Q\left(\lfloor \gamma_0 + 1\rfloor,  \left(\frac{P_s}{ \pi\rho^2} + \lambda_n \right) A T  \right) \right)\sum_{i=1}^{k+1} \int_{\Rc_{i-1}}^{\Rc_i} \frac{r}{\sigma^2}e^{-\frac{r^2}{2\sigma^2}}\, dr      \left( Q\left(\lfloor \gamma_0 + 1\rfloor,  \left(\frac{P_s}{ \pi\rho^2} + \lambda_n \right) A T  \right) \right)^{k-i} ,
\end{align}
for $k = 0, 1, \dotsc, N-1.$  Also, 
\begin{align}
p = \sum_{k=0}^{N-1} \int_{\Rc_k}^{\Rc_{k+1}}  \frac{r}{\sigma^2} e^{-\frac{r^2}{2\sigma^2}}\, dr \times \left(Q\left(\lfloor \gamma_0 + 1\rfloor,  \left(\frac{P_s}{ \pi\rho^2} + \lambda_n \right) A T  \right) \right)^{N-k}.
\end{align}
Finally, it can be shown that $\sum_{k=1}^N P(E_k) + p = 1.$
\begin{figure}
    \centering
    \includegraphics{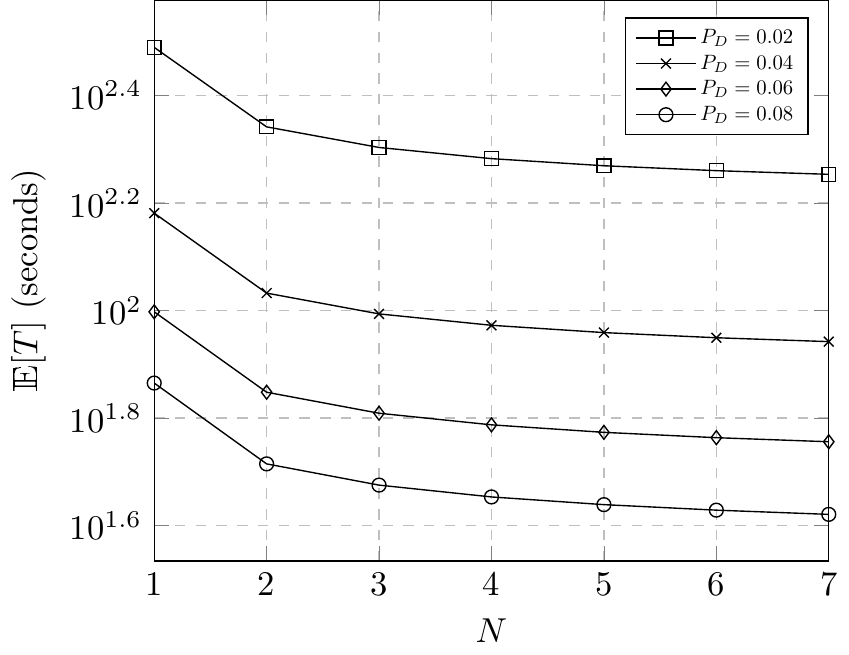}
    \caption{Plots of average acquisition time as a function of number of subregions $N$ when the radii $\Rc_i$'s are uniformly distributed between 0 and $\Rc$ for different values of probability of detection $P_D$. The radius of uncertainty region $\Rc = 50$ meters, the standard deviation of the receiver location inside the uncertainty region $\sigma = 15$ meters, beam radius $\rho=0.2$ meters, and dwell time $T_d = 0.1$ millisecond.}
    \label{fig1}
\end{figure}

\begin{figure}
    \centering
    \includegraphics{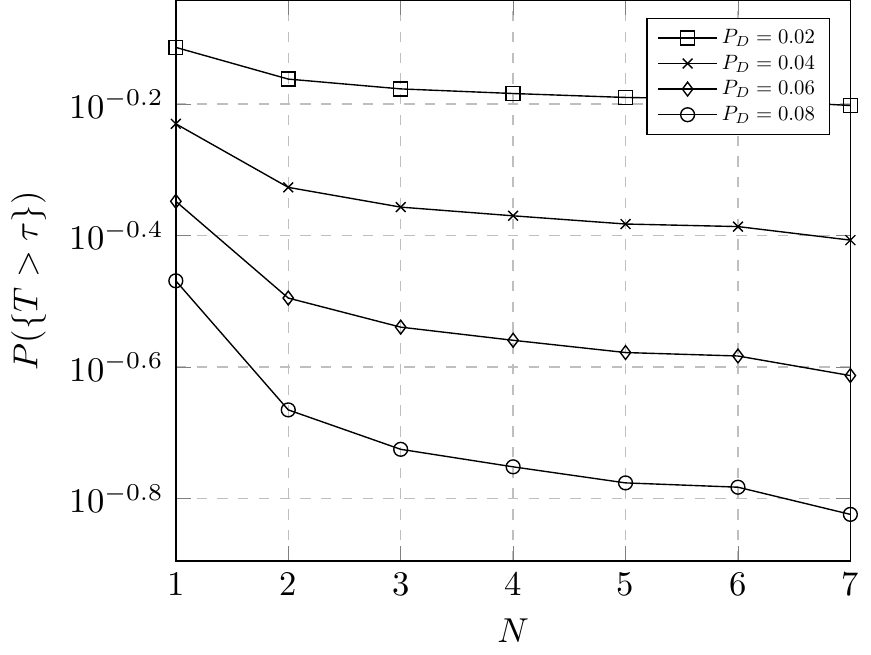}
    \caption{Plots of complementary cumulative distribution function $P(\{ T > \tau \})$ as a function of number of subregions $N$ when the radii $\Rc_i$'s are uniformly distributed between 0 and $\Rc$ for different values of probability of detection $P_D$. The radius of uncertainty region $\Rc = 50$ meters, the standard deviation of the receiver position inside the uncertainty region $\sigma = 15$ meters, beam radius $\rho=0.2$ meters, dwell time $T_d = 0.1$ millisecond, and the time threshold $\tau = 80$ seconds. }
    \label{fig2}
\end{figure}

\begin{figure}
	\centering
	\includegraphics{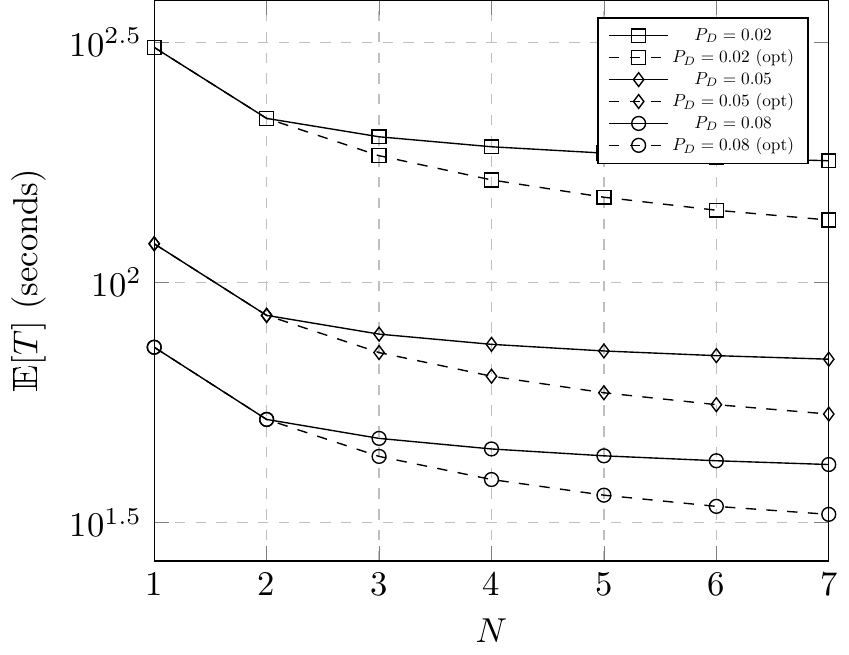}
	\caption{Plots of mean acquisition time as a function of number of subregions $N$ for the optimized and nonoptimized cases. The radius of uncertainty region $\Rc = 50$ meters, the standard deviation of the receiver position inside the uncertainty region $\sigma = 15$ meters, beam radius $\rho=0.2$ meters, and dwell time $T_d = 0.1$ millisecond. }
	\label{fig3}
\end{figure}

\begin{figure}
	\centering
	\includegraphics{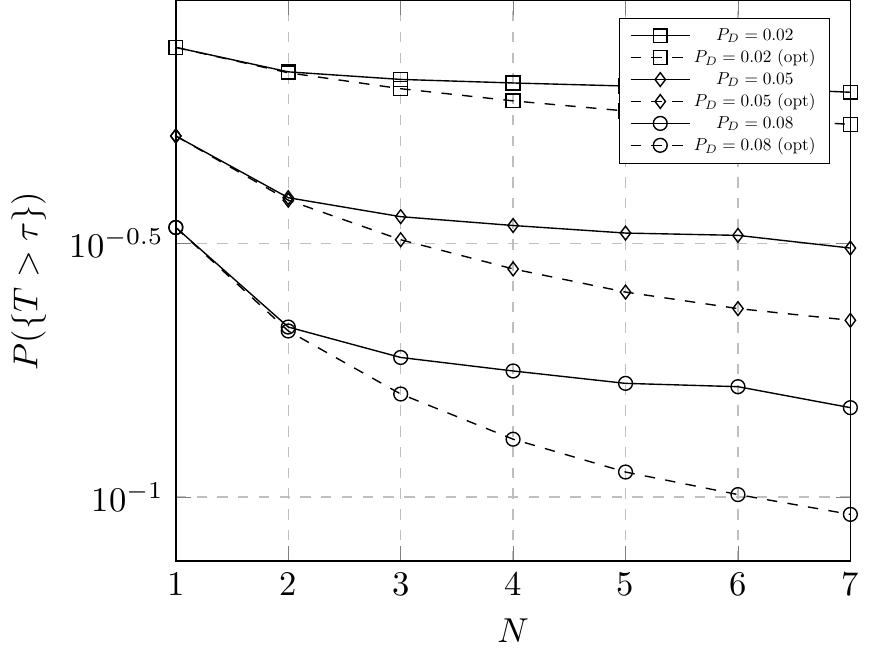}
	\caption{Plots of complementary cumulative distribution function $P(\{ T > \tau \})$ as a function of number of subregions $N$ for the optimized and nonoptimized cases. The radius of uncertainty region $\Rc = 50$ meters, the standard deviation of the receiver position inside the uncertainty region $\sigma = 15$ meters, beam radius $\rho=0.2$ meters, dwell time $T_d = 0.1$ millisecond, and time threshold $\tau = 80$ seconds. }
	\label{fig4}
\end{figure}

\subsubsection{Complementary Cumulative Distribution Function}

\begin{align}
    &P(\{ T > \tau \}) = \int_{\tau}^\infty f_T(t)\, dt = \sum_{i=0}^\infty p^i \sum_{k=0}^{N-1} P(E_{k+1}) \int_{\tau}^\infty \frac{\alpha}{\eta_{k+1}} e^{-\alpha\left(t - i \beta_N - \beta_k \right)}\cdot \mathbbm{1}_{[i\beta_N + \beta_k, i\beta_N + \beta_{k+1} )}(t)\, dt. 
\end{align} 
In terms of indicator functions, the complementary distribution function can be written more simply as 
\begin{align}
 &P(\{ T > \tau \})= \sum_{i=0}^\infty p^i  \sum_{k=0}^{N-1} \! P(E_{k+1})  \left( \mathbbm{1}_{[0, i\beta_N + \beta_k )}(\tau) + \mathbbm{1}_{[i\beta_N + \beta_k, i\beta_N + \beta_{k+1} )}(\tau) \frac{1}{\eta_{k+1}} \left(e^{-\alpha \left( \tau - i \beta_N - \beta_k \right) } - e^{-\alpha \left( \beta_{k+1} - \beta_k \right) }\right) \right).
 \end{align}
 After a few manipulations, we arrive at the following closed form expression of the complementary distribution function:
 \begin{align}
 P(\{ T > \tau \})&=  \sum_{k=0}^{N-1}P(E_{k+1}) \left( \sum_{i = \max\left(0, \left\lfloor \frac{\tau-\beta_k}{\beta_N} \right\rfloor + 1 \right)}^\infty p^i + \sum_{i = \max\left(0, \left\lfloor \frac{\tau-\beta_{k+1}}{\beta_N} \right\rfloor + 1 \right)}^{\left\lfloor \frac{\tau-\beta_k}{\beta_N} \right\rfloor} \frac{1}{\eta_{k+1}} \left( p e^{\alpha \beta_N} \right)^i e^{-\alpha \left( \tau - \beta_k \right)} \right. \nonumber \\
 & \left. - \sum_{i = \max\left(0, \left\lfloor \frac{\tau-\beta_{k+1}}{\beta_N} \right\rfloor + 1 \right)}^{\left\lfloor \frac{\tau-\beta_k}{\beta_N} \right\rfloor} \frac{1}{\eta_{k+1}} p^i e^{-\alpha \left( \beta_{k+1} - \beta_k \right) } \right) \nonumber \\
 & = \sum_{k=0}^{N-1}P(E_{k+1}) \left( \frac{p^{\max\left(0, \left\lfloor \frac{\tau-\beta_k}{\beta_N} \right\rfloor + 1 \right)}}{1-p} + \frac{e^{-\alpha(\tau - \beta_k)}}{\eta_{k+1}} \times \left(p e^{\alpha \beta_N}\right)^{\max\left(0, \left\lfloor \frac{\tau-\beta_{k+1}}{\beta_N} \right\rfloor + 1 \right)} \times \frac{1 - \left(p e^{\alpha \beta_N}\right)^{L_k}}{1 - p e^{\alpha \beta_N}} \right.\nonumber \\
 & \left. - \frac{e^{-\alpha(\beta_{k+1}-\beta_k)}}{\eta_{k+1}} \times p^{\max\left(0, \left\lfloor  \frac{\tau-\beta_{k+1}}{\beta_N} \right\rfloor + 1 \right)} \times \frac{1 - p^{L_k}}{1 - p}  \right),
\end{align}
where $L_k \coloneqq \left\lfloor \frac{\tau - \beta_k}{\beta_N}  \right\rfloor - \max\left(0, \left\lfloor \frac{\tau-\beta_{k+1}}{\beta_N} \right\rfloor + 1 \right) + 1$.

\begin{figure}
	\centering
	\includegraphics{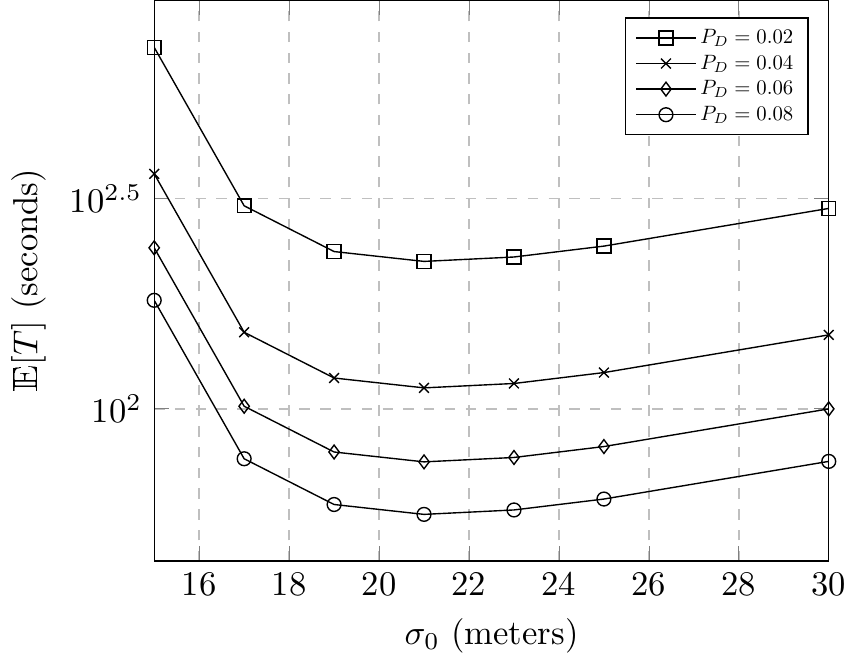}
	\caption{Plots of average acquisition time as a function of standard deviation of firing distribution  $\sigma_0$ for the shotgun approach. The radius of uncertainty region $\Rc = 50$ meters, the standard deviation of the receiver position inside the uncertainty region $\sigma = 15$ meters, beam radius $\rho=0.2$ meters, and dwell time $T_d = 0.1$ millisecond. The optimal value of $\sigma_0$ is $\sqrt{2}\sigma = 21.2132$ meters.}
	\label{fig8}
\end{figure}

\begin{figure}
	\centering
	\begin{tabular}{ll}
		\includegraphics[scale=0.9]{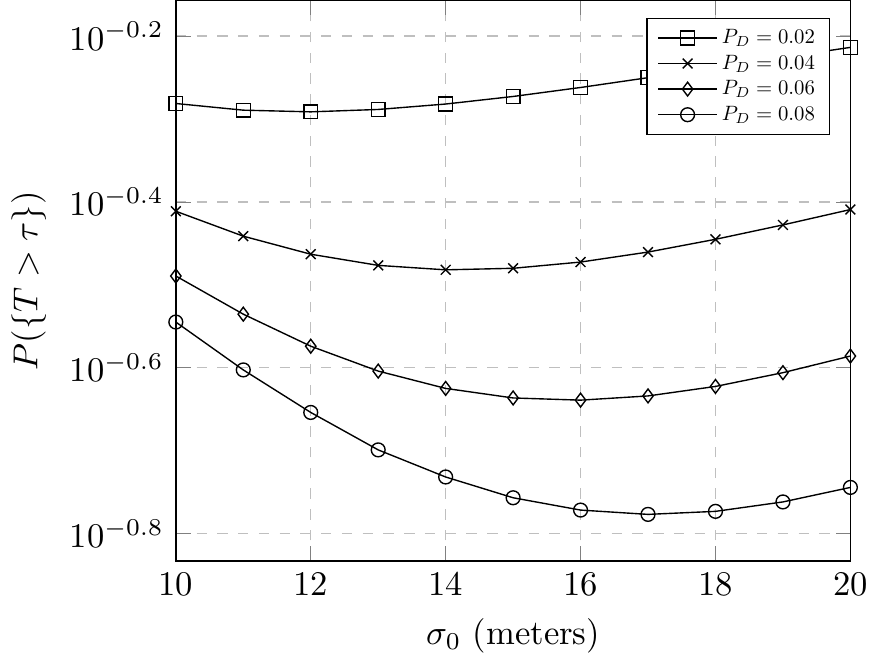}
		&
		\includegraphics[scale=0.9]{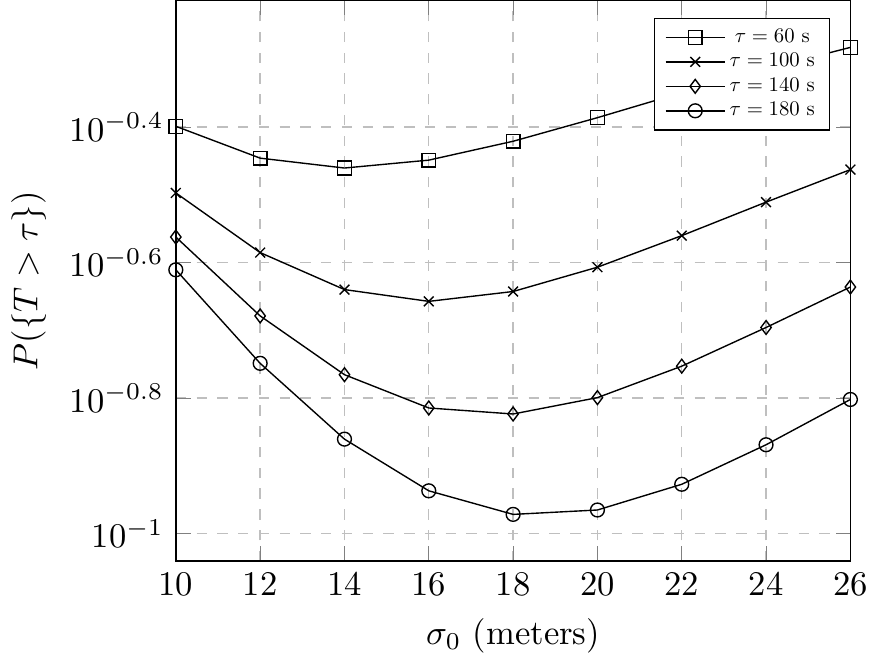}
	\end{tabular}
	\caption{The plots of complementary cumulative distribution function as a function of standard deviation of firing distribution $\sigma_0$. Left: Plot of $P(\{ T > \tau \})$ for time threshold $\tau = 80$ seconds for different values of probability of detection $P_D$.  
		Right: Plot of $P(\{ T > \tau \})$ for probability of detection $P_D = 0.05$ for different values of time threshold $\tau$. The radius of the uncertainty region $\Rc = 50$ meters, the standard deviation of the receiver position inside the uncertainty region  $\sigma = 15$ meters, beam radius $\rho=0.2$ meters, and dwell time $T_d = 0.1$ milliseconds for both the figures.}
	\label{fig10}
\end{figure}

\begin{figure}
	\centering
	\includegraphics{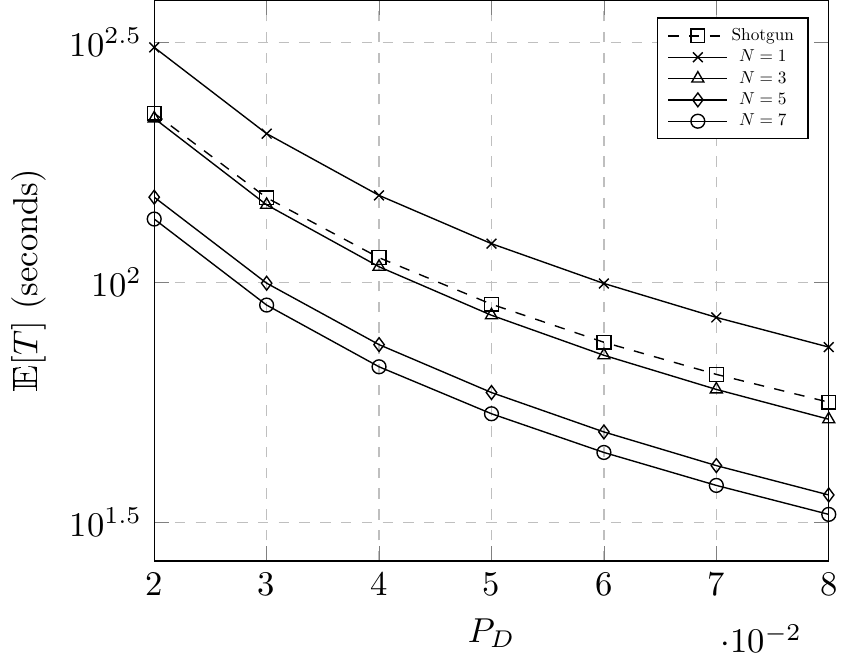}
	\caption{Plots of average acquisition time as a function of probability of detection $P_D$ for the optimized shotgun and adaptive spiral schemes. The radius of the uncertainty region $\Rc = 50$ meters, the standard deviation of the receiver position inside the uncertainty region $\sigma = 15$ meters, beam radius $\rho=0.2$ meters, and dwell time $T_d = 0.1$ millisecond. }
	\label{fig5}
\end{figure}

\begin{figure}
	\centering
	\includegraphics{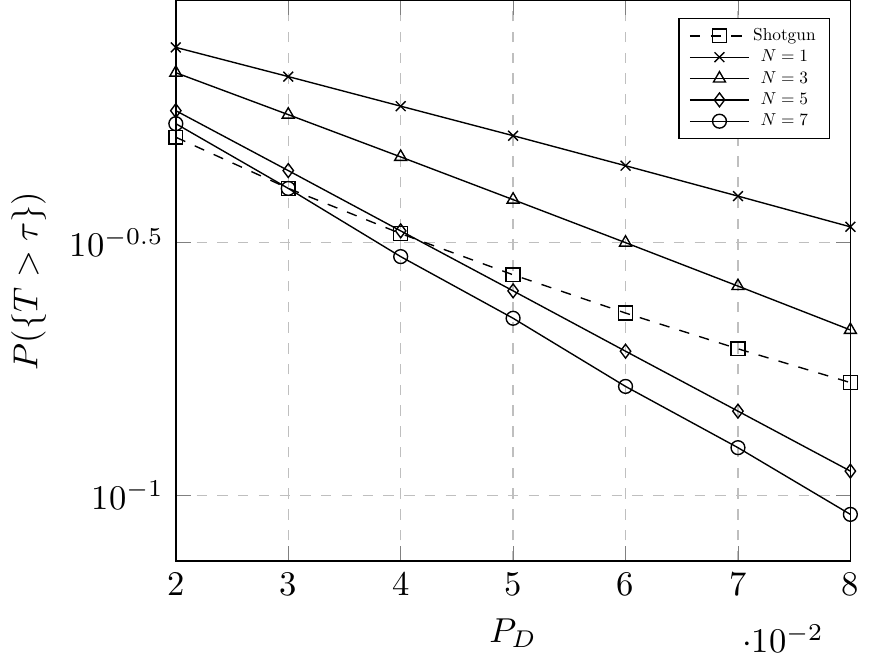}
	\caption{Plots of complementary cumulative distribution function $P(\{ T > \tau \})$ as a function of probability of detection $P_D$ for the optimized shotgun and adaptive spiral schemes. The radius of the uncertainty region $\Rc = 50$ meters, the standard deviation of the receiver position inside the uncertainty region $\sigma = 15$ meters, beam radius $\rho=0.2$ meters, dwell time $T_d = 0.1$ millisecond, and time threshold $\tau = 80$ seconds. }
	\label{fig6}
\end{figure}

\begin{figure}
	\centering
	\includegraphics[scale=1.5]{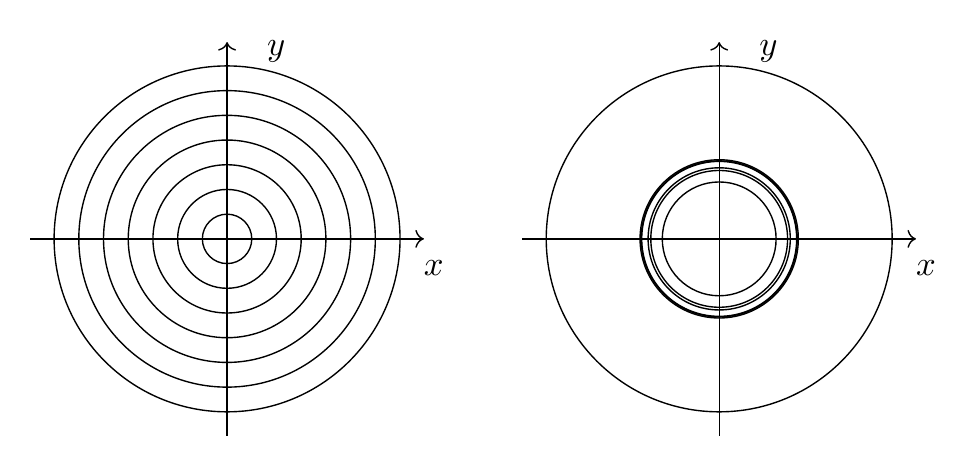}
	\caption{Distribution of regions for the adaptive spiral search scheme before and after optimization of mean acquisition time when the uncertainty region is divided into $N=7$ subregions. For probability of detection $P_D = 0.05$, the unoptimized and optimized schemes yield mean acquisition times of 69.19 and 53.27 seconds, respectively. }
	\label{fig7}
\end{figure}

 \subsection{Optimization of $\E[T]$ and $P(\{T > \tau\})$}
 In this section, we look to optimize $\E[T]$ as a function of $\Rc_1, \dotsc, \Rc_{N-1}$
 when $\rho$ is fixed. The optimization problem for $\E[T]$ and $P(\{ T > \tau \})$ is laid out as follows:
 \begin{equation*}
\begin{aligned}
& \underset{\Rc_1, \dotsc, \Rc_{N-1}}{\text{minimize}}
& & f(\Rc_1, \dotsc, \Rc_N)  \\
& \text{subject to}
& & i) 0<\Rc_1 < \Rc_2 < \dotsm < \Rc_{N-1} < \Rc_N , \: i = 1, 2, \dotsc, N,\\
&&& ii) \rho = \rho_0,\\
& & & iii) P_R = P_0,
\end{aligned}
\end{equation*}
where $f(\Rc_1, \dotsc, \Rc_N)$ is either $\E[T]$ or $P(\{ T > \tau\})$,  $P_R$ is the received signal power, and$\rho_0$ and $P_0$ are constants.

The optimization is not performed as a function of $\rho$, and the smallest possible value of $\rho$ (which is $\rho_0$) is chosen for scanning. This is because enlarging $\rho$ results in a further decrease in an already small probability of detection $P_D$, and instead of saving time by scanning with a larger beam radius, a larger time is incurred whenever $\rho > \rho_0$ (due to a poorer $P_D$).

Finally, we want to point out that the solution of the optimization problem with the method of Lagrange multipliers is not straightforward in our case. Therefore, for the purpose of global optimization, we use a \emph{real-number genetic algorithm} to find the minimum of the objective function. For more details about the real-number genetic algorithm, the reader is referred to \cite{Rao}.


\section{The Shotgun Approach}\label{SG}
The so-called \emph{Shotgun} approach to acquisition is a randomised acquisition technique that involves firing the uncertainty region with signal pulses at certain locations in the region, where the locations are sampled according to a zero-mean Gaussian distribution in two dimensions. We term such a Gaussian distribution as the \emph{firing} distribution. In the remaining section, we derive the mean acquisition time and the complementary cumulative distribution function of the acquisition time for the shotgun approach.

Let $\B$ be the event that the beam falls inside a ball of radius $\rho$ that contains the receiver. For the sake of analysis, we first assume that the receiver is located at a point $(x, y)$ inside the uncertainty region. Let $B_\rho(x, y)$ be such a ball of radius $\rho$ centered around $(x, y)$. We assume that when the beam center falls inside $B_\rho(x, y)$, the detector is completely covered by the beam, and there is a chance of detection. In this case, the probability of occurrence of $\B$, given that the receiver is located at $(x, y)$, is
\begin{align}
    P(\B|x, y) = \iint_{B_\rho(x, y)} \frac{1}{2\pi \sigma_0^2} e^{-\left(\frac{x^2+y^2}{2\sigma_0^2} \right)} \, dx\, dy, \label{shot1}
\end{align}
where $\sigma_0^2$ is the variance of the firing distribution. For the practical case of $\rho << \sigma_0$, the expression in \eqref{shot1} is merely, 
\begin{align}
    P(\B|x,y) \approx  \frac{\pi \rho^2}{2\pi \sigma_0^2} e^{-\left(\frac{x^2+y^2}{2\sigma_0^2} \right)} =  \frac{ \rho^2}{2 \sigma_0^2} e^{-\left(\frac{x^2+y^2}{2\sigma_0^2} \right)}.
\end{align}
Let us use $p_D(x, y)$ to denote the probability of detection of the receiver when one shot is fired in the uncertainty region. Then, 
\begin{align}
    p_D(x, y) = P(\B|x, y) P_D.
\end{align}
In this case, the acquisition time, $T$, has the (geometric) distribution:
\begin{align}
    f_T(t|x, y) \coloneqq \sum_{\ell = 1}^\infty (1-p_D(x, y))^{\ell-1} p_D(x, y) \delta(t - \ell T_d),
\end{align}
which implies that $\E[T|x, y] = \frac{T_d}{p_D(x, y)}$.
\subsection{Mean Acqusition Time}
The mean acquisition time can be shown to be
\begin{align}
    \E[T]& = \int_{-\infty}^\infty \int_{-\infty}^\infty \E[T|x, y] f_{X Y}(x, y) \, dx \, dy = \int_{-\infty}^\infty \int_{-\infty}^\infty \frac{T_d}{p_D(x, y)} \frac{1}{2\pi \sigma^2} e^{-\left(\frac{x^2+y^2}{2\sigma^2} \right)}\, dx \, dy\nonumber \\
    & = \frac{2 T_d \sigma_0^4}{\rho^2 P_D (\sigma_0^2 - \sigma^2)}, \text{ for } \sigma_0 > \sigma. \label{shot2}
\end{align}
The proof of \eqref{shot2} is given in the Appendix~\ref{C}.

The mean acquisition time is optimized (minimized) with respect to $\sigma_0$. By taking the partial derivative of \eqref{shot2} with respect to $\sigma_0$, and setting the resulting derivative equal to zero, we obtain
\begin{align}
    \sigma_0^* = \sqrt{2} \sigma.
\end{align}

\subsection{Complementary Cumulative Distribution Function}
The conditional complementary cumulative distribution function of $T$ is derived as follows:
\begin{align}
&P(\{ T > \tau \}|x, y) = \int_\tau^\infty f_T(t|x, y) \, dt =    \sum_{\ell = 1}^\infty (1-p_D(x, y))^{\ell-1} p_D(x, y) \int_\tau^\infty \delta(t - \ell T_d) \, dt\nonumber \\
& = \sum_{\ell = 1}^\infty (1-p_D(x, y))^{\ell-1} p_D(x, y)  \cdot \mathbbm{1}_{[0, \ell T_d)}(\tau) = \frac{p_D(x, y)}{(1-p_D(x, y))}  \sum_{\ell = \max\left(1, \left\lfloor \frac{\tau}{T_d} \right\rfloor + 1 \right)}^\infty (1-p_D(x, y))^{\ell} \nonumber \\
& = \left(1-p_D(x, y)\right)^{\max\left(1, \left\lfloor \frac{\tau}{T_d} \right\rfloor + 1 \right)  - 1}. \label{ccdf}
\end{align}
Let us express the conditional complementary distribution in an alternative form that will help us obtain a closed-form expression when we average it with respect to the receiver location. Before we start the analysis, let us denote the factor $\max\left(1, \left\lfloor \frac{\tau}{T_d} \right\rfloor + 1 \right)  - 1$ by the integer $n$. We can rewrite \eqref{ccdf} as
\begin{align}
&P(\{ T > \tau \}|x, y) = \left(1-p_D(x, y)\right)^n = \sum_{k=0}^{n} {{n} \choose {k} } (-1)^k (p_D(x, y))^k \nonumber \\
& = \sum_{k=0}^{n} {{n} \choose {k} } (-P_D)^k \left(\frac{ \rho^2}{2 \sigma_0^2} e^{-\left(\frac{x^2+y^2}{2\sigma_0^2} \right)}\right)^k = \sum_{k=0}^{n} {{n} \choose {k} } (-P_D)^k \left(\frac{ \rho^2}{2 \sigma_0^2}\right)^k \exp\left( {-\frac{x^2+y^2}{2 \frac{\sigma_0^2}{k}  } } \right)
\end{align}
Thus,
\begin{align}
P(\{ T > \tau \}) &= \int_{-\infty}^\infty \int_{-\infty}^\infty P(\{ T > \tau \}|x, y) f_{XY}(x, y)\, dx \, dy \nonumber \\
& = \int_{-\infty}^\infty \int_{-\infty}^\infty  \sum_{k=0}^{n} {{n} \choose {k} } (-P_D)^k \left(\frac{ \rho^2}{2 \sigma_0^2}\right)^k \exp\left( {-\frac{x^2+y^2}{2 \frac{\sigma_0^2}{k}  } } \right) \frac{1}{2 \pi \sigma^2} \exp\left(  - \frac{x^2+y^2}{2\sigma^2} \right) \, dx \, dy \nonumber \\
& = \sum_{k=0}^{n} {{n} \choose {k} }  \left(\frac{-P_D \rho^2}{2 \sigma_0^2}\right)^k \frac{1}{2\pi \sigma^2} \int_{-\infty}^\infty \int_{-\infty}^\infty \exp\left(  - \frac{(x^2 + y^2)}{2} \left( \frac{1}{\sigma^2} + \frac{k}{\sigma_0^2} \right) \right) \, dx \, dy \nonumber \\
& = \sum_{k=0}^{n} {{n} \choose {k} }  \left(\frac{-P_D \rho^2}{2 \sigma_0^2}\right)^k \left(\frac{\sigma_0^2}{\sigma_0^2 + k \sigma^2} \right)  \int_{-\infty}^\infty \int_{-\infty}^\infty \frac{1}{2\pi \left( \frac{\sigma^2 \sigma_0^2}{\sigma_0^2 + k \sigma^2}  \right) } \exp\left(  - \frac{(x^2 + y^2)}{2 \left( \frac{\sigma^2 \sigma_0^2}{\sigma_0^2 + k \sigma^2}\right)}  \right)  \, dx \, dy \nonumber \\
& = \sum_{k=0}^{n} {{n} \choose {k} }  \left(\frac{-P_D \rho^2}{2 }\right)^k \left(\frac{1}{\sigma_0^2 + k \sigma^2} \right) \left( \frac{1}{\sigma_0^2} \right)^{k-1}, \label{ccdf1}
\end{align}
where, we note that $n = \max\left(1, \left\lfloor \frac{\tau}{T_d} \right\rfloor + 1 \right)  - 1 = \max\left( 0, \left\lfloor \frac{\tau}{T_d} \right\rfloor \right)$. For a small $T_d$, $n$ can be a very large number, and it become very difficult to calculate \eqref{ccdf1} due to the factor ${n \choose k}$ (which is not easy to calculate when $n$ is large and $k$ is moderately large, $k < n$). However, all the three terms in the sum in \eqref{ccdf1} approach zero as $k >> 1$. Therefore, we don't have to compute the entire sum in \eqref{ccdf1} because the terms in the sum beyond some integer $n_0$ can be ignored, where $n_0 << n$. Thus, with $n_0$ as the upper limit in the sum, the complementary cumulative distribution can be computed easily with a small approximation error.

The optimization of the complementary cumulative distribution function is carried out by differentiating \eqref{ccdf1} with respect to $\sigma_0$ and setting it equal to zero. However, the solution (which is $\sigma_0^*$, the minimizer), has to be computed numerically. We ought to note here that $\sigma_0^*$ is a function of both $\tau$ and $P_D$.

\section{Performance Results and Comparison of Two Acquisition Schemes} \label{Comp}

In order to compare the performance of the adaptive acquisition schemes, the number of signal photons detected during the observation interval is fixed at 25, and the number of noise photons is varied between 13 and 24. These photon counts result from the following system parameters: Received signal intensity $ = 6\times 10^{-8}$ Joules/square meters/second,  average noise intensity $= 4\times 10^{-8}$ Joules/square meters/second, area of detector $ = 1$ square centimeter, wavelength of light $ =1550$ nanometers, pulse duration $=1$ microsecond, and photoconversion efficiency $=0.5$. By using these parameter values in \eqref{PD1} and \eqref{PFA}
 and choosing an appropriate threshold, we get the  probability of detection to lie between 0.02 and 0.08, while the probability of false alarm is fixed at $1\times 10^{-12}$. 

Fig.~\ref{fig1} and Fig.~\ref{fig2} show the expected acquisition time and the complementary cumulative distribution function (ccdf) of acquisition time, respectively, as a function of number of subregions $N$ for the adaptive spiral search scheme. These curves correspond to the uniform spacing between the radii $\Rc_i$'s which is the nonoptimized scenario. We can see that the acquisition performance  improves with $N$. However, as can be studied from these curves, the law of diminishing return takes effect when $N$ grows. Fig.~\ref{fig3} and Fig.~\ref{fig4} depict the gain in performance achieved by optimization of the radii of the subregions. As can be seen, the performance gains can be significant when $N$ is large. Fig.~\ref{fig7} shows the distribution of the radii $\Rc_i$'s for the optimized versus  nonoptimized  scenarios. Here, we remind the reader that the  $N=1$ case corresponds to the regular spiral search that is employed in \cite{XinLi} and \cite{Bashir6}.

It should be noted that we cannot choose $N$ arbitrarily large since the optimization of the radii $\Rc_i$ becomes computationally expensive for a large number of radii. As can be inferred from Fig.~\ref{fig3} and Fig.~\ref{fig4}, the performance gains with optimization of acquisition time for a smaller $N$ yields better results as compared to nonoptimized case when $N$ is large. The maximum value we chose for $N$ is 7 in this study for both the optimized and nonoptimized scenarios. 

Fig.~\ref{fig8} and Fig.~\ref{fig10} illustrate the mean acquisition time and the complementary cumulative distribution function, respectively, for the shotgun approach as a function of standard deviation of the firing distribution.  We note that the optimal value $\sigma_0^*$ is a function of $P_D$ as well as $\tau$ in the case of the complementary cumulative distribution function, whereas, $\sigma_0^*$ is independent of $P_D$ for the mean acquisition time scenario (Fig.~\ref{fig8}). Fig.~\ref{fig5} and Fig.~\ref{fig6} represent the difference in performance between the shotgun approach and the adaptive spiral scheme as a function of $P_D$. Both schemes are optimized to give the best possible performance. As can be seen, the shotgun approach gives a better performance than the $N=1$ and $N=2$ scenarios from the perspective of complementary cumulative distribution function, but is outperformed by larger $N$ for the adaptive spiral search for higher $P_D$.

Even though the shotgun approach does not perform as well as the adaptive spiral search for a larger $N$, this approach is still desirable from two important perspectives. Let us remember that in the spiral acquisition, we have to trace the spiral carefully while scanning the uncertainty region. This requires a very high pointing accuracy on the transmitter's part. In a real system, their is always a pointing error tolerance limit within which the transmitter system operates, and if the magnitude in error is significant, the performance of the adaptive spiral search can take a serious hit. Intuitively, we can see why this is true. If the transmitter misses the receiver due to the pointing error, it will have to scan an entire subregion before it gets a chance to shine light again on the receiver. On the other hand, the pointing error is not such a serious problem for the shotgun approach since the pointing error only results in slightly increasing the uncertainty volume (assuming that the GPS localization error and the transmitter's pointing error are independent random variables). 

In addition to a need for higher pointing accuracy, the optimization cost (cost of executing a real-number genetic algorithm in a multidimensional space) of adaptive spiral search may also make it a less suitable choice. On the other hand, the optimization of mean acquisition time as a function of $\sigma_0$ is very easy to carry out for the shotgun approach. However, the task of optimization of  complementary cumulative distribution function for the shotgun scheme may be more computationally intensive.

\section{Conclusion}\label{Conc}
In this paper, we have proposed and analyzed two acquisition schemes, namely the adaptive spiral search, and the shotgun approach. In terms of acquisition time, both schemes perform better than the regular spiral search scheme for low probability of detection scenario. For a large number of subregions, the adaptive spiral search outperforms the shotgun technique. However, in order to gain better performance, the adaptive search spiral requires precise pointing by the transmitter in order to scan the region of uncertainty. Additionally, the optimization of adaptive spiral search using a genetic algorithm may also incur additional complexity overhead. This may tilt the balance in favor of shotgun approach which does not require higher pointing accuracy and larger optimization cost.

\appendices

\section{Probability of Receiver Location inside $E_{S_i}$}\label{A}
Since, $\left(\bigcup_{j=1}^{k-1} E_j \right)^C = \bigcap_{j=1}^{k-1} E_j^C$,
\begin{align}
P\left(E_{S_i} \left|  \bigcap_{j=1}^{k-1} E_j^C \right. \right) =
\int_0^{\Rc_i} f_R\left(r\left|  \bigcap_{j=1}^{k-1} E_j^C \right. \right) \, dr.
\end{align}
Now, 
\begin{align}
f_R\left(r\left|  \bigcap_{j=1}^{k-1} E_j^C \right. \right) = \frac{ P\left(\bigcap_{j=1}^{k-1} E_j^C| R=r\ \right) f_R(r)} {P\left(\bigcap_{j=1}^{k-1} E_j^C \right)}. \label{1}
\end{align}
Moreover, we know that,
\begin{align}
P\left(\bigcap_{j=1}^{k-1} E_j^C| R=r\ \right) = \begin{dcases}
1, &  r > \Rc_{k-1},\\
1-P_D, &  \Rc_{k-2} < r \leq \Rc_{k-1}, \\  
1 - (P_D + (1-P_D)P_D), & \Rc_{k-3} < r \leq \Rc_{k-2},\\
\vdots & \vdots \\
1 - (P_D + (1-P_D)P_D + (1-P_D)^2 P_D + \dotsm + (1-P_D)^{k-2} P_D), & \Rc_0 < r \leq \Rc_1,
\end{dcases}
\end{align}
where $R_0 = 0.$ We can see that if $P_D \approx 0 \implies P\left(\bigcap_{j=1}^{k-1} E_j^C| R=r\ \right) \approx 1$ for any $r$. Using this result in \eqref{1}, we conclude that for $P_D \approx 0$, $f_R\left(r\left|  \bigcap_{j=1}^{k-1} E_j^C \right. \right) \approx f_R(r)$ since $P\left(\bigcap_{j=1}^{k-1} E_j^C \right)$ is not a function of $r$, and acts only as a normalization constant. This shows that 
\begin{align}
P\left(E_{S_i} \left|  \bigcap_{j=1}^{k-1} E_j^C \right. \right) =
\int_0^{\Rc_i} f_R\left(r\left|  \bigcap_{j=1}^{k-1} E_j^C \right. \right) \, dr \approx \int_0^{\Rc_i} f_R\left(r \right) \, dr = P(E_{S_i}).
\end{align}

\section{Conditional Distribution of $X$} \label{B}
 By Bayes Rule, the conditional distribution of the radius of the receiver location in the uncertainty region is:
\begin{align}
f_X(t|E_k) = \frac{P(E_k |t) f_X(t)}{P(E_k)}. \label{2}
\end{align}
Moreover, $P(E_k|t) = 0$ for $r > T_d\Rc_k^2/\rho^2,$ and 
\begin{align}
P(E_k | t)=(1-P_D)^{k-i}P_D, \quad T_d\Rc_{i-1}^2/\rho^2 < t \leq T_d \Rc_i^2/\rho^2, i \leq k.
\end{align}

 For $P_D\approx 0$, $P(E_k|T)\approx P_D$ for  $ 0< t \leq T_d\Rc_k^2/\rho^2$. Thus,  \eqref{2} can be approximated as
\begin{align}
f_X(t|E_k) = \frac{P_D}{P(E_k)} f_X(t) \cdot \mathbbm{1}_{ (0,  T_d\Rc_k^2/\rho^2]}(t),
\end{align}
where $f_X(t) = \alpha \exp(-\alpha t) \cdot \mathbbm{1}_{[0, \infty)}(t),$ and $\frac{P_D}{P(E_k)}$ acts as a normalization constant.

\section{Mean Acquisition Time for Shotgun Approach} \label{C}
From \eqref{shot2}, we have that
\begin{align}
\E[X] &= \int_{-\infty}^\infty \int_{-\infty}^\infty \frac{T_d}{p_D(x, y)} \frac{1}{2\pi \sigma^2} e^{-\left(\frac{x^2+y^2}{2\sigma^2} \right)}\, dx \, dy = \int_{-\infty}^\infty \int_{-\infty}^\infty \frac{T_d}{\frac{1}{2\pi \sigma_0^2}\exp\left(\frac{ -(x^2+y^2)}{2\sigma_0^2} \right) \pi \rho^2 P_D} \times \frac{1}{2\pi \sigma^2} e^{-\left(\frac{x^2+y^2}{2\sigma^2} \right)}\, dx \, dy\nonumber \\
& = \frac{2 T_d \sigma_0^2}{\rho^2 P_D} \int_{-\infty}^\infty \frac{1}{2\pi \sigma^2}e^{-\left(\frac{x^2+y^2}{2\sigma^2} \right)} e^{\left(\frac{x^2+y^2}{2\sigma_0^2} \right)} \, dx \, dy = \frac{2 T_d \sigma_0^2}{\rho^2 P_D} \int_{-\infty}^\infty \frac{1}{2\pi \sigma^2}e^{-\left(\frac{(x^2+y^2)(\sigma_0^2-\sigma^2)}{2\sigma_0^2 \sigma^2} \right)}  \, dx \, dy\nonumber \\
&= \frac{2 T_d \sigma_0^2}{\sigma^2 \rho^2 P_D}\times \frac{\sigma_0^2 \sigma^2}{(\sigma_0^2-\sigma^2)}  \int_{-\infty}^\infty \frac{1}{2\pi \frac{\sigma_0^2 \sigma^2}{(\sigma_0^2-\sigma^2)}  }\exp\left({-\frac{(x^2+y^2)}{ 2\frac{\sigma_0^2 \sigma^2}{(\sigma_0^2 - \sigma^2)} } }\right)  \, dx \, dy \nonumber \\
& = \frac{2 T_d \sigma_0^4}{\rho^2 P_D (\sigma_0^2 - \sigma^2)}.
\end{align}

\bibliography{reference1}
\bibliographystyle{IEEEtran}
\end{document}